\documentclass[12pt]{article}
\usepackage{amsmath}
\usepackage{graphicx,psfrag,epsf}
\usepackage{enumerate}
\usepackage{natbib}
\usepackage{url} 

\usepackage{amsmath,amssymb}
\usepackage{graphicx}
\usepackage{vmargin,bbm,times,graphicx,amssymb}

\usepackage{hyperref, natbib, float}
\usepackage{lineno}
\usepackage[margin=1cm]{caption}

\newcommand{\blind}{0}

\begin{document}

\def\spacingset#1{\renewcommand{\baselinestretch}%
{#1}\small\normalsize} \spacingset{1}


\if0\blind
{
  \title{\bf Uncertainty in phylogenetic tree estimates}
  \author{Amy Willis 
    \hspace{.2cm}\\
    Department of Biostatistics,  University of Washington\\\
    and \\
    Rayna Bell \\
    Smithsonian Institution, National Museum of Natural History}
  \maketitle
} \fi

\if1\blind
{
  \bigskip
  \bigskip
  \bigskip
  \begin{center}
    {\LARGE\bf Title}
\end{center}
  \medskip
} \fi

\bigskip
\begin{abstract}
Estimating phylogenetic trees is an important problem in evolutionary biology, environmental policy and medicine. Although trees are estimated, their uncertainties are generally discarded in statistical models for tree-valued data. Here we explicitly model the multivariate uncertainty of tree estimates. We consider both the cases where uncertainty information arises extrinsically (through covariate information) and intrinsically (through the tree estimates themselves). The latter case  is applicable to any procedure for tree estimation, and thus has broad relevance to the entire field of phylogenetics. The importance of accounting for tree uncertainty in tree space is demonstrated in two case studies. In the first instance, differences between gene trees are small relative to their uncertainties, while in the second, the differences are relatively large. Our main goal is visualization of tree uncertainty, and we demonstrate advantages of our method with respect to reproducibility, speed and preservation of topological differences compared to visualization based on multidimensional scaling. The proposal highlights that phylogenetic trees are estimated in an extremely high-dimensional space, resulting in uncertainty information that cannot be discarded. Most importantly, it is a method that allows biologists to diagnose whether differences between gene trees are biologically meaningful, or due to uncertainty in estimation.
\end{abstract}

\noindent%
\vfill

\newpage
\spacingset{1.45} 

\section{Introduction} \label{intro}

Information about the shared evolutionary history of organisms  is critical in many different contexts. A correct understanding of evolutionary relationships and  the mechanisms by which lineages diverge informs conservation efforts \citep{marko2011genetic}, environmental policy \citep{carpenter2010comparative}, and medicine \citep{gonzalez2013molecular}. Consequently, the field of phylogenetics, which concerns the inference of evolutionary relationships between organisms, is an active and well-developed subdiscipline of biology.

Collections of phylogenetic trees are common in phylogenetic inference and typically consist of individual phylogenetic trees each constructed from different parts of the genome, or of a subset of phylogenetic trees sampled from the posterior distribution of a phylogenetic analysis. Effective graphical representations of trees can quickly indicate whether a collection of trees represent similar evolutionary histories; however, visualizing collections of trees is challenging because phylogenetic trees are multivariate, tree-valued data objects.

Almost all phylogenetic trees depicted in the modern biological literature convey uncertainty by reporting the proportion of tree samples (bootstrap or posterior) on which the branch is present. However, this approach treats each branch as independent, ignoring the multivariate nature of trees. Here we propose a procedure that accounts for dependencies in the trees' branches, and models  their  uncertainty multivariately. Our proposal uses the log map of \cite{blow2} to map trees from their  metric space to Euclidean space, where we model a tree estimate $\hat{T}$ as a noisy realization of the true tree $T$.  Once we estimate the model parameters, we can then employ Euclidean multivariate analysis techniques to reduce the dimension of the trees,  and visualize them, along with their uncertainties.

While one motivation for this paper is to propose a new method for modeling and visualizing uncertainty in trees, another motivating factor concerns the treatment of trees  by mathematicians working with tree space. The metric space of phylogenetic trees was developed to assist with tree comparisons, tree estimation  and quantification of  uncertainty  in trees  \citep{bhv}. While substantial progress has been made towards the first two  of these goals \citep{feragen2012hierarchical, Chakerian:2012kj, bbb, dinh2016consistency}, incorporating tree uncertainty into {\it statistical} models (which we distinguish from recent advances in pure probability) has been relatively untreated in the literature. Trees in tree space are almost always viewed as point masses, which falsely suggests that they are known with certainty. While flexible tree estimation methods are available \citep{mrbayes,beast2,Mirarab:2014kw}, phylogenetic trees are nonetheless estimated, and information regarding estimation precision is generally available. This paper advocates for the inclusion of this precision when considering trees as objects in their metric space.

The paper begins by introducing phylogenetic trees along with a discussion of existing tree visualization methods, including multidimensional scaling (Section \ref{procedure}). The log map is then proposed as a visualization tool for trees and tree uncertainty (Section \ref{logmap}). This is followed by a discussion of some sources of uncertainty in phylogenetic trees, and some methods for incorporating univariate uncertainty into tree visualization and modeling using the log map (Section \ref{extrinsic_s}). Multivariate uncertainty is then introduced (Section \ref{intrinsic}), and the advantages of the procedure described here are contrasted with multidimensional scaling (Section \ref{contrast}). The paper is concluded with some limitations of the log map (Section \ref{limitations})  and some final remarks (Section \ref{conclusion}). All scripts and data are available via \href{https://github.com/adw96/TreeUncertainty}{github}.

\section{Tree visualization} \label{procedure}

Phylogenetic trees are edge-weighted trees (acyclic graphs with no nodes of degree 2). They represent the evolutionary relationships of a collection of organisms that are known to share an ancestor. Internal vertices (vertices/nodes with degree 3 or greater) represent the shared ancestor that existed  at a point in history, and the leaves (also called tips; vertices with  degree one) represent modern (or observable, eg. via the fossil record) organisms. The length of the branches (the edge-weights) represents the extent of divergence  between nodes. In some contexts the graph will have a ``root'' representing the common ancestor. In these cases the graph will be directed and the root will have out-degree one.

The space of possible trees is enormous.  To illustrate, consider that a tree with $m+2$ leaves has at most $m$  internal branches of positive length, and there are  $\frac{(2m+2)!}{2^{m+1}(m+1)!}$ possible tree topologies. By comparison, there are $2^m$  orthants in $\mathbbm{R}^m$. Since each tree can be associated with a single orthant in $\mathbbm{R}^m$ (that reflects its branch lengths), it can thus be argued that tree space is $$\frac{(2m+2)!}{2^{m+1}(m+1)!} \bigg/ 2^m = 
\frac{(2m+2) \times (2m+1) \times \ldots \times (m+2)}{2^{2m+1}}$$ times larger than Euclidean space; that is, exponentially larger. Thus the space of trees with 6 leaves  is  more than 50  times larger than $\mathbbm{R}^4$, and the space of trees with 10 leaves  is   more than 130,000  times larger than $\mathbbm{R}^8$. As a result, representing collections of trees, especially large collections of trees sharing a large leaf set, necessitate some compression of the trees' information.

The particular form of tree space compression that we advocate in this paper is called the log map. The log map is a function that maps from tree space to Euclidean space, and was first proposed by \cite{blow2}. It captures both topology (branching order) and branch length information about a tree. Here we present the log map as a modeling tool for both trees and tree uncertainty, arguing for its advantages as a native coordinate system for trees, in reproducibility of tree analyses, and for its capacity to reflect multivariate uncertainty in tree estimates.

Models and algorithms for estimating phylogenetic trees abound in the literature \citep{mrbayes,beast2,Mirarab:2014kw, binet2016fast}. In this section we do not consider the problem of tree estimation. Instead, we wish to visualize a collection of trees that share a leaf set of $m$ taxa. These trees may arise via different models for tree estimation, as samples from a posterior distribution for a given gene's phylogeny, or as estimates of the phylogeny based on different genes. We denote them as $\hat{T}_1, \ldots, \hat{T}_n$, permitting them to be either rooted or unrooted. We wish to visualize this collection.

\subsection{Multidimensional scaling} \label{mds}

A common method of visualizing collections of trees is multidimensional scaling \citep{hillis2005analysis,Chakerian:2012kj, Kendall24062016}. Multidimensional scaling (MDS), first proposed by \cite{torgerson1952multidimensional}, is a powerful technique for mapping a collection of objects to vectors in $\mathbbm{R}^k$, where $k$ is usually chosen to be 2 or 3. The only requirement is a distance (or dissimilarity measure) between the objects. MDS involves finding a matrix that minimizes a stress function, which encodes the difference between the distances under the map  and the true distances. The minimising matrix, in $\mathbbm{R}^{n \times k}$, can then be visualized as $n$ points in $\mathbbm{R}^k$. A common choice of stress function is the Kruskal-1 function \citep{kruskal1964multidimensional, hillis2005analysis}, with the resulting map given by
\begin{align}
\arg \min_{x \in \mathbbm{R}^{n \times k}} \left( \sum_{i \neq j} D_{ij}-|x_i-x_j|^2 \right)^{1 / 2}, \label{mds-equation}
\end{align}
where $\{D_{ij}\}_{(i,j) \in \{1, \ldots, n\}^2} = d(\hat{T}_i, \hat{T}_j)$ is the $n \times n$ matrix of distances between the tree-valued estimates $\hat{T}_i, \hat{T}_j$.

A key advantage of MDS is that the distance can be chosen to best highlight the differences of importance between the trees in the collection. \cite{hillis2005analysis} proposed using the Robinson-Foulds (RF) distance \citep{robinson1981comparison} to view the topological differences between trees. \cite{Chakerian:2012kj} discussed the advantages of using the BHV distance to incorporate both topological and branch length features, and  \cite{Gori:2016vr}  used this approach to cluster genes by phylogeny. \cite{Kendall24062016} recently proposed a method for comparing differences between the most recent common ancestors of the tips, and \cite{TreeScaper} describe an efficient implementation of MDS for many different tree metrics.  We will contrast the advantages of MDS with the advantages of our procedure in Section \ref{contrast}.

\subsection{DensiTree}

The program DensiTree \citep{bouckaert2010densitree, bouckaert2014densitree} is a popular program for viewing collections of trees, and integrates seamlessly with the most prominent Bayesian tree estimation programs. DensiTree overlays the trees transparently, thus darker regions of the image imply greater confidence. Furthermore, small numbers of alternative topologies with comparable levels of support are easily observed. This tool can also show other parameters that are used in coalescent-based phylogenetic tree estimation, such as population size, by indexing the widths of the branches to these parameters. DensiTree has the advantage that the graphical representation shows the collection of trees as trees, rather than mapping the trees to Euclidean space. It effectively illustrates both topological and branch length disagreements, but performs most effectively when only a small number of topologies conflict and the tree has  a relatively small number of leaves. Large leaf sets, or many conflicting topologies, are difficult to distinguish by eye. The new procedure that we propose here can clearly show conflicting clusters of trees, and works well when there are many such clusters. It is at the expense, however, of visualizing the tree directly.

\subsection{Other tree visualization methods}

Other methods for comparing phylogenetic trees have been proposed in the literature, and we briefly mention some that are designed for comparing more than two trees (we will not address pairwise comparisons given that our interest is in comparing large collections of trees, nor will we discuss methods for viewing individual trees). \cite{Sundberg:2009hy} argue that the space of resolved trees can be mapped to an $n$-torus, and then uses cartographic projections to reduce dimension. However, this loses key branch length information and non-resolved trees are not permitted. TreeScaper \citep{TreeScaper} implements various tools for analyzing collections of trees, including estimating intrinsic dimensionality, and community detection. A dimension reduction tool that removes selected branches to aid longitudinal analysis has recently been proposed \citep{zairis2016genomic}, though the authors' goal was not visualization. Heatmaps can be used to summarize the dissimilarity matrix of MDS \citep{puigbo2007topd}. Treemaps, which partition a rectangle into panels representing each tree and then display the nodes in a space-filling manner, may  be used to see hierarchical dissimilarities \citep{tu2007visualizing}, though its scalability with the size of the collection is limited. Trees of trees, proposed by \cite{nye_trees_of_trees}, assigns the topologies of the trees in the collection to a ``meta-tree'''s leaf nodes, then uses neighbour-joining methods to cluster the nodes (trees) by topolgical similarity.  \cite{Hess:2014wb} append histograms of related parameter estimates to leaves to show variation in these parameters (eg. population size) across the clades, while  \cite{Bremm:2011hx} developed  PhyloComp to enable targeted comparison of differences at recent or deep divergence times (note the connection to the motivation of the metric proposed by \cite{Kendall24062016}). treespace \citep{treespace}, implemented as an R package, is a  modern tool approach to analyzing distributions of trees, which is known to be a difficult problem. Most of these methods are designed to highlight topological differences, and few scale well both visually and computationally.

\section{The log map} \label{logmap} \label{overview}

Each of the methods described above have distinct advantages in different situations. However, none have the ability to illustrate  uncertainties in the $\{\hat{T}_i\}_i$'s. The method that we propose here, which deals with this issue,  utilizes a map from tree space to Euclidean space, called the log map. We review the construction of the log map here.

Consider the trees $T_1, \ldots, T_n$ with the same leaf set as objects in $\mathcal{T}_{m+3}$, the metric space of phylogenetic trees \citep{bhv}. While tree space has many locally Euclidean properties, tree topologies are globally folded over each other, and thus  visualizing the space directly is extremely difficult. Furthermore, because tree space is not equipped with an inner product, we have no concept of orthogonality or rotations on the space, which prohibited descriptions of covariance and central limit theorems until recently. To deal with the latter, \cite{blow2} developed the log map, $\Phi_{T^*}(T): \mathcal{T}_{m+3} \rightarrow \mathbbm{R}^m$, which gives the  Euclidean coordinates of a target tree $T$ relative to a base tree $T^*$.  The following 4 cases are helpful in understanding the log map:
\begin{enumerate}
\item $\Phi_{T^*}(T^*)$: The log map returns an arbitrary ordering of the $m$ branch lengths of $T^*$ (some of which may be zero for an unresolved tree $T^*$).
\item $\Phi_{T^*}(T)$ for $T$ with the same topology as $T^*$: The log map returns the branch lengths of $T$ in accordance with the ordering of $\Phi_{T^*}(T^*)$.
\item $\Phi_{T^*}(T)$ for $T$ with a topology that is a single nearest neighbor interchange (NNI) from $T^*$: The same as for Case 2 for all shared edges present on both trees, and the coordinate of the branch that is not shared is $-e_{T},$ where $e_T$ is the length of the branch present on $T$ but not on $T^*$.
\item $\Phi_{T^*}(T)$  for $T$ with a topology that is two or more NNIs from $T^*$: Let $\gamma(T^*, T)$ be the geodesic distance between $T^*$ and $T$. Then $\Phi_{T^*}(T)$ is found by continuing for distance $\gamma(T^*, T)$ along a unit vector in the direction from $T^*$ towards the first bend in the BHV geodesic. See \cite{Willis:UOx4Gul3} or  \cite{blow2} for a more formal description.
\end{enumerate}

We refer the reader to \cite{blow2} for a discussion of the case where $T^*$ falls on codimensional planes in $\mathcal{T}_{m+3}$.

In many ways, the log map is an ideal map because it preserves both local directions and BHV distances between trees (with respect to the base tree $T^*$). Some information must be discarded in any mapping from tree space to Euclidean space because tree space cannot be completely embedded in Euclidean space (e.g., \citet[p. 2720]{nye11}). However, while the topological information of trees is lost, the information as to whether or not a tree was one or more NNIs from the base tree is maintained: one NNI is given by a single negative coordinate, more than one NNI is denoted by multiple negative coordinates. We return to a discussion of this in Section \ref{contrast}  and Figure \ref{mammals_branches}. The log map fulfills the suggestion of \cite{Holmes:2005wu} to map trees to $\mathbbm{R}^d$, though the construction is very different to that of MDS.

\section{Modeling tree uncertainty using extrinsic information} \label{extrinsic_s}

The log map preserves information with respect to both tree topology and branch lengths, which suggests its potential utility for both modeling and visualizing multivariate tree uncertainty. In this section we consider a model that incorporates covariate information about the precision in the tree   estimates, which we call {\it extrinsic information} because it is not tree-valued. We first describe the model and then demonstrate it on an example.

\subsection{Model setup} \label{extrinsic}

Consider  trees $\hat{T}_1, \ldots, \hat{T}_n$: $n$ estimates of an underlying evolutionary history $T$. Suppose extrinsic, that is, non-tree, information is available regarding the precision of the estimates. The example that we will consider  is information regarding the evolutionary rate of each of the genes whose gene trees are estimated by the $\hat{T}_i$'s. However, the Fr\'echet variance \citep{bbb, Benner:2014uy}, a parsimony or likelihood score \citep{montealegre2002visualizing}, or a measure of the extent of taxon sampling could also be used. The setting is that each estimate $\hat{T}_i$ is associated with a known covariate $r_i$, which is proportional to the variance of the estimate in a manner we now define.

We begin by projecting the estimates to Euclidean space using the log map. The log map requires a base tree on which to ground the projection, and here we use the proposal of \cite{Willis:UOx4Gul3} to ground the log map at the sample Fr\'echet mean of the estimates. However, we are now in the heteroscedastic case, and need to modify this accordingly. Thus we define our base tree as  the weighted Fr\'echet mean
\begin{align}
\overline{{T}} := \arg \min_{t \in \mathcal{T}_m} \sum_{i=1}^n \frac{1}{r_i} \gamma(\hat{T}_i, t)^2,  \label{f_mean2}
\end{align} where $\gamma(t_1, t_2)$ is the distance between trees $t_1$ and $t_2$ in the BHV metric. Note that the CAT(0) property of the BHV  space guarantees tree mean uniqueness \cite[Theorem 2.4]{bacak}, which is an additional reason for choosing it here.


We now suppose that the log maps of the estimates are independent, noisy realisations of the log map of the true tree, that is,
\begin{align}
\Phi_{\overline{T}} (\hat{T}_i) = \Phi_T(T) + \mathcal{N}(0, \sigma^2 r_iI_m), \label{extrinsic_model}
\end{align}
where $T$  is the population analogue of  $\overline{T}$:
\begin{align}
T := \arg \min_{t \in \mathcal{T}_m} \int_{\mathcal{T}_P} \frac{ \gamma(T, t)^2}{r_{{T}}} dF(T),  \label{f_mean}
\end{align}
for $F$ the tree-generating process (e.g. the population dynamics process that drives speciation across the population of genes). This model is underpinned by the belief that all estimates reflect the underlying phylogenetic tree, but that their precision differs in accordance with the extrinsic information known about the tree. Both the log map of $T$ and the common variance parameter $\sigma ^ 2$ can then be estimated using maximum likelihood:
\begin{align}
\widehat{ \Phi_T(T)} &= \frac{\sum_i \Phi_{\overline{T}} (\hat{T}_i)/r_i}{\sum_i 1/r_i}, \label{logmapest} \\
\hat{\sigma}^2 &= \frac{1}{n} \sum_i \frac{(\Phi_{\overline{T}} (\hat{T}_i)-\widehat{ \Phi_T(T)} )^2}{r_i}. \label{sigsqest}
\end{align}

We are now able to associate the estimates with their information content visually. Consider  a set with measure $(1-\alpha)$  under a normal distribution with mean $\Phi_{\overline{T}} (\hat{T}_i)$ and variance $\hat{\sigma}^2r_iI_m$ (we suggest the minimum volume set, which is centred at $\Phi_{\overline{T}} (\hat{T}_i)$). This set exists in $\mathbbm{R}^m,$ and thus to assist visualization we suggest projecting it onto the span of the first two principal components of the $\{\hat{T}_i\}$ and visualizing it on this subspace. Note that these sets are not confidence nor  prediction sets, but are useful analogues for viewing the variability in the tree estimates.

Invariably objections could be raised as to the choice of the noise model, which we modelled as a spherical multivariate normal distribution. Because we only have a single realisation of each tree, for identifiability  our covariance matrices must be low rank. We consider using multiple tree estimates in Section \ref{intrinsic}  to relax this constraint. However, Gaussian distributional cross-sections may not be plausible (though note that no literature regarding the distribution of log map estimates has yet been developed). If a practitioner wished to relax the assumption of normality, the only necessary changes to the procedure would be recalculating the maximum likelihood estimates of Equations (\ref{logmapest}) and (\ref{sigsqest}), and changing the contours of the visualization set. While model misspecification should be investigated and corrected as just described,  the purpose of this procedure is visualization, not inference, and we argue that this objective can be achieved despite mild deviations from normality. See \cite{feragen2012hierarchical} and \cite{amenta2015quantification} for two-sample permutation-based tests for equality of tree means and variances, and \cite{Holmes:2005wu} for a more general discussion of tree testing.

We now demonstrate this approach in a situation where the evolutionary rate of genes is available. We conclude the discussion of extrinsic tree uncertainty modeling by noting that if multivariate information regarding tree uncertainty was available ($\{r_i^{(1)}, \ldots, r_i^{(\ell)}\}_i),$ it could be incorporated into the above procedure by modeling the perturbations by $\mathcal{N}(0,\sigma_{(1)}^2r_i^{(1)} + \ldots + \sigma_{(\ell)}^2r_i^{(\ell)}),$ subject to identifiability constraints (no linear dependence amongst the vectors ($\{r_i^{(1)}, \ldots, r_i^{(\ell)}\}_i).$

\subsection{Evolutionary rate and phylogenetic uncertainty} \label{univariate}

Comparing sets of phylogenetic trees, each estimated using different regions of the genome (e.g. loci), is complicated by variation in the evolutionary rates of those regions. Loci that evolve slowly contain few informative sites for estimating recent divergence events, while loci that evolve more quickly often contain more natural variation, or mutational saturation, than phylogenetic signal for resolving older divergence events \citep{wagele2007visualizing, baeza2013exploring}. The informativeness of a particular locus for resolving a particular tree (i.e. ancient versus recent divergence) is thus related to its evolutionary rate.

To illustrate how the visualization method described in Section \ref{extrinsic} can be used to incorporate uncertainty from covariate information, we consider the relative evolutionary rate of 574 different genes shared by 42 mammals. The OrthoMam database \citep{ranwez2007orthomam, douzery2014orthomam} contains estimates of the gene trees of these genes, which we call $\hat{T}_1, \ldots, \hat{T}_{574}$ (estimation details available in \cite{ranwez2007orthomam}; the most commonly-selected model of best fit was $\text{GTR}+\Gamma$), along with the rates, which we call $r_1, \ldots, r_{574}$. Rate estimation was performed by \cite{ranwez2007orthomam}  via the Super Distance Matrix procedure of \cite{adkins2001molecular}, which involves finding the rescaling of branches necessary to equalize leaf-to-tip distances.

\begin{figure}
\begin{center}
\includegraphics[trim = 0cm 0.5cm 0.7cm 0.5cm, scale=1]{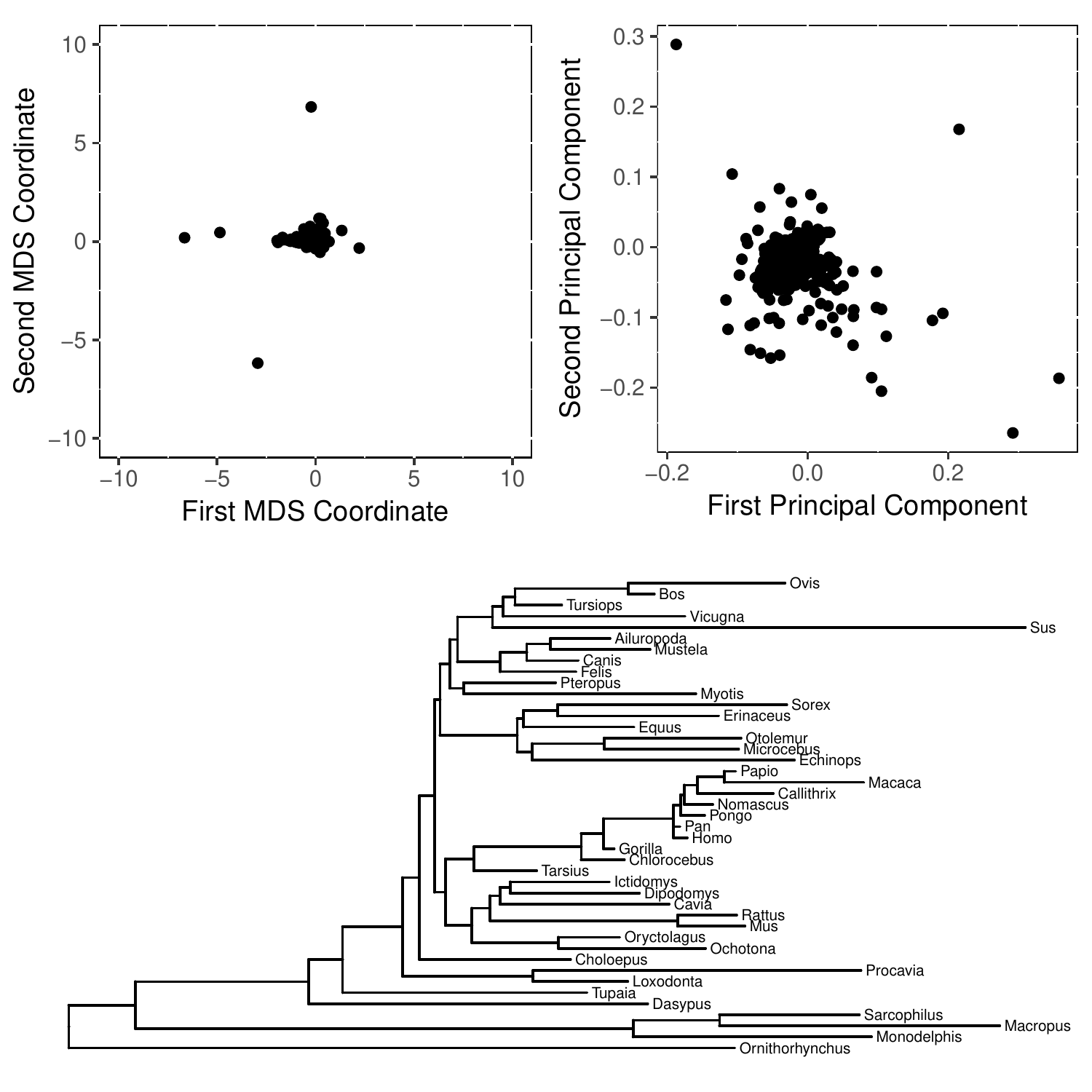}
\caption{574 gene trees shared by 42 mammals  \citep{douzery2014orthomam}. (top left) Multidimensional scaling of the BHV distances between the trees. (top right) The first two principal components of the log map of the trees with respect to their weighted Fr\'echet mean. Note that both representations suggest that the trees are known rather than estimated. (bottom) The SLC7A2 gene tree, one of the 574 gene trees mapped by the top figures. }
\label{orthomam1}
\end{center}
\end{figure}

Multidimensional scaling of the trees with respect to the BHV  distance \citep{Chakerian:2012kj} is shown in Figure \ref{orthomam1} (top left), along with the first two principal components of the log maps of the trees (top right).  The first 2 principal components explain 64\% of the variation in the 40-dimensional log mapped trees (see Supplementary Data for projections onto the third and fourth principal components). Both representations suggest that there are  4  trees that differ from the rest. MDS emphasizes this more  heavily, most likely because the objective of MDS  is to best capture {\it all} of the pairwise distances in the mapping, while the log map instead preserves distances to the base tree $\overline{T}$. 
These 4  trees are topologically distinct from $\bar{T}$ by multiple nearest neighbor interchanges (see Supplementary Data).

\begin{figure}
\begin{center}
\includegraphics[trim = 0cm 0.5cm 0.7cm 0.5cm, scale=0.8]{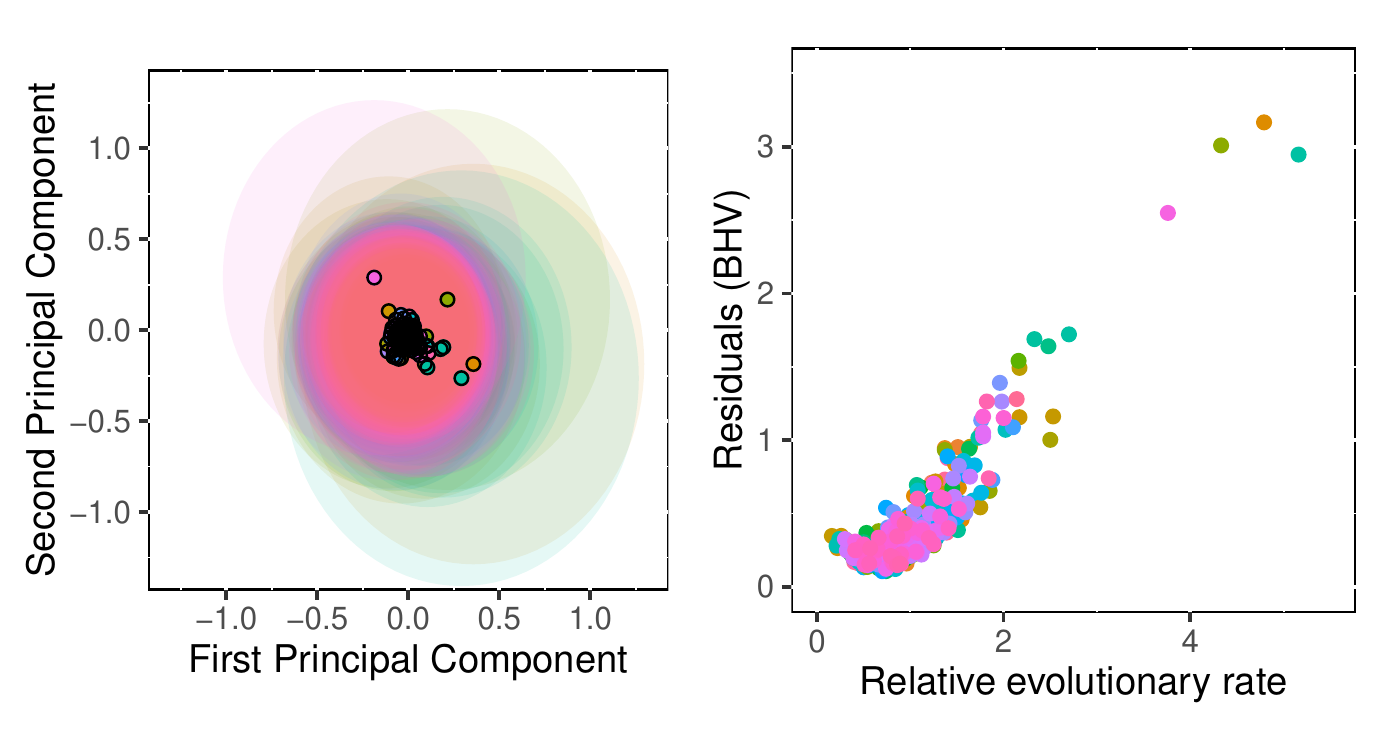}
\caption{To account for estimation error, we construct a model for tree uncertainty (Section \ref{extrinsic}). (left) The 40-dimensional sets of volume 0.95 representing each tree, projected onto the first two principal components of the log maps of the tree estimates. Note that the sets appear large because they were constructed in a much higher dimensional  space, and not because of the 4 ``outliers'' (see text). (right) The BHV distance from each gene tree to the weighted Fr\'echet mean shown against the relative evolutionary rates of the genes. Lower evolutionary rates, but not the lowest, correspond to the minimum distance trees.}
\label{orthomam2}
\end{center}
\end{figure}

We begin by calculating the weighted Fr\'echet mean of the gene trees $\overline{T}$ with weights $1/r_i$ (Algorithm 4.2 of \cite{bacak} setting $\lambda_k = 1/k$; see also \cite{mopr}), and finding the log maps of the gene trees with respect to $\overline{T}$, $\Phi_{\overline{T}}(\hat{T}_i)$. We then calculate the maximum likelihood estimates of the model parameters (Equations (\ref{logmapest}) and (\ref{sigsqest})), and  construct 0.95-measure sets with respect to a $\mathcal{N}(\Phi_{\overline{T}}(\hat{T}_i), \hat{\sigma}^2r_iI_{40})$ distribution to illustrate the disagreement between the trees {\it after accounting for the variance model}. Because these sets are in $\mathbbm{R}^{40}$, we project them onto the first two principal components\footnote{Note that constructing the sets and then projecting them is computationally wasteful. For this reason we only determine the algebraic form of the $\mathbbm{R}^{40}$ sets, and determine the form of the $\mathbbm{R}^{2}$ sets algebraically before constructing them. Details  are given as Supplementary Materials.} of the $\{\Phi_{\overline{T}}(\hat{T}_i)\}_i.$ These projected sets are shown in Figure \ref{orthomam2} (left).

The impression given by  Figure \ref{orthomam1} suggests that there are a small number of trees  that are very different from the rest of the collection. However, Figure \ref{orthomam2} (left) makes it clear that relative to the uncertainty in the tree estimates, these trees are not especially outlying. A natural question to ask is to what extent the ``outliers'' inflate the estimate of $\sigma^2$, and removing these 4 points and recalculating the variance reduces it by only 5.4\%. Alas, the variance of the collection is only one  component of the size of the sets; the other is the dimension. These log maps have been estimated in a 40-dimensional space, and the radius of these sets grows with the square root of the dimension. We argue that the high dimensionality of tree estimation leads to the uncertainty in estimation that we observe in Figure \ref{orthomam2} (left). Note that while the sets in $\mathbbm{R}^{40}$ sets are spherical, the rotation that we use is not, and for this reason the sets do not appear isotropic.

It is interesting to note the relationship between the distance between $\overline{T}$ and the $\Phi_{\overline{T}}(\hat{T}_i)$'s and the $r_i$. While the form of Equation (\ref{f_mean2}) places the greatest weight on gene trees with low evolutionary rate, we see  from Figure \ref{orthomam2} (right) that these are not the closest gene trees to the weighted average tree: trees corresponding to genes with small but non-minimal  evolutionary rates are the minimum distance trees.  We thus conjecture that the weighted Fr\'echet mean may provide a good estimate of the overall evolutionary process by incorporating the evolutionary rate information into tree estimation.

\section{Intrinsic uncertainty information} \label{intrinsic}

Having discussed the case where covariate information is available regarding the tree estimates, we now consider the case where the trees themselves can be used for inferring the precision in estimation.  We call this intrinsic information because it is tree-valued. In particular, suppose we wish to estimate $k$ different phylogenies $T^{(1)}, \ldots, T^{(k)}.$ After describing the model in Section \ref{intrinsic1}, we consider the case where these are the phylogenies of different genes in Section \ref{multivariate}.

\subsection{Intrinsic model setup} \label{intrinsic1}

The intrinsic uncertainty information comes from having $n_i$ estimates of $T^{(i)}$, which we call $\hat{T}_1^{(i)}, \ldots, \hat{T}_{n_i}^{(i)}.$ Clearly this collection (which would most commonly arise from bootstrapping or posterior sampling) contains multivariate information about tree uncertainty: branches that can be estimated precisely would be present on all trees and with low variance in their lengths, while contentious branches may only appear on some trees or have huge variation in their lengths relative to the mean length. We wish to visualize these estimates and their multivariate uncertainty.

Define $\overline{T}^{(i)}$ to be the unweighted Fr\'echet mean of the $\{\hat{T}^{(i)}_j\}_j$'s, $\overline{T}$ to be the unweighted Fr\'echet mean of the $\overline{T}^{(i)}$, and $T$ to be the population analogue of $\overline{T}$. We note that the latter may not have any biological significance, but we construct it order to have a common base for the log map. 
We now consider the model  $$\Phi_{\overline{T}} (\hat{T}_j^{(i)}) = \Phi_T(T^{(i)}) + \mathcal{N}(0, \Sigma_i).$$ The key difference between this model and that of Section \ref{extrinsic} is that the uncertainty of the estimates is not constrained to be spherical: we permit $\Sigma_i$ to be unstructured. However, we can use the collection $\{\Phi_{\overline{T}}(\hat{T}_j^{(i)})\}_{j}$ to estimate it. We suggest estimation via maximum likelihood if $n_i$ is large relative to  the dimension of tree space, or a structured estimator if not.

Similar to the proposal of Section \ref{extrinsic}, we construct $(1-\alpha)$-volume sets of the distribution of the $\Phi_{\overline{T}} (\hat{T}_j^{(i)}),$ and again suggest the minimum volume sets. Again the sets will be $m$-dimensional, and we suggest projecting them to the subspace in $\mathbbm{R}^2$ that spans the first two principal components of the $\{\Phi_{\overline{T}}(\hat{T}_j^{(i)})\}_{i,j}$.

A key element of this visualization strategy is that it maintains $m$-dimensional uncertainty of tree estimation. Constructing a set based on a model for the principal components gives the impression of substantially more precision than truly exists, and MDS prohibits meaningful model constructions because the coordinate system is sample-dependent (discussed in Section \ref{contrast}). Visualizing uncertainty in tree estimates is an extremely important issue because of documented overconfidence in phylogeny estimates. Strong support for conflicting topologies can arise among phylogenies constructed from different sets of loci (e.g. sequence capture versus restriction site associated DNA sequencing [RADseq] \citep{leache2015phylogenomics}) or from the same dataset analyzed under different models (e.g. concatenation versus multispecies coalescent \citep{edwards2016implementing}). We hope our procedure assists with visualizing the uncertainty of phylogenetic estimates under these different scenarios.

\subsection{Multivariate tree uncertainty} \label{multivariate}

Here we will consider using samples from a posterior distribution on tree space to generate collections of tree estimates. However, the procedure proposed in Section \ref{intrinsic1} is agnostic with respect to the origin of the estimates.\footnote{ Substituting posterior standard deviations for frequentist standard errors overstates the precision in the estimates when priors are chosen to be uninformative \citep{Efron:1996vx, efron2015frequentist}, as is often the case in phylogenetic inference. The recent proposal of \cite{efron2015frequentist} could be applied to tree space in order to correct for this. However, because we are focusing only on an exploratory method, we defer the generalization of Efron's proposal to tree space to a separate investigation.} Bootstrap resampling is another plausible method for generating collections of tree estimates.

Bayesian methods for estimating phylogenies naturally give rise to collections of trees as samples from the posterior. While sophisticated methods for summarizing samples from a posterior distribution on tree space exist (Section \ref{procedure}), tools to visualize the information that is lost in summarizing the collection of trees by a single tree are relatively underdeveloped.

To illustrate the advantages of the  proposal of Section \ref{intrinsic1} to utilize collections of tree estimates to capture the multivariate tree uncertainty, we consider visualizing discordance between mitochondrial and nuclear gene phylogenies.
In general, phylogenetic inferences based on different genes for a given set of taxa may differ with respect to topology and/or relative branch length due to poor gene tree reconstruction (e.g. due to mutational saturation and homoplasy) or because the gene trees differ from the underlying species tree (e.g. due to incomplete lineage sorting or introgression \citep{Maddison:1997de, coalescent2}).  In particular, mitochondrial DNA can be especially problematic for resolving phylogenetic relationships at deeper evolutionary timescales due to its higher evolutionary rate of change.
A reasonable visualization procedure should showcase the relative uncertainties in inferring the phylogenies of different loci.


\cite{wiens2010discordant} and \cite{spinks2009conflicting} investigated discordance between mitochondrial and nuclear loci in reconstructing evolutionary relationships among species of Emydid turtles. The large number of splits on the resulting tree estimates challenges fast identification of the differences between the mtDNA and nuDNA phylogeny estimates (see \citet[Figures 1 and 2]{wiens2010discordant}). We selected a subset of 10 species that form a monophyletic group in the combined mtDNA and nuDNA analysis in \cite{wiens2010discordant} and combined data from \cite{wiens2010discordant}, \cite{spinks2009conflicting}  and \cite{angielczyk2011adaptive} to create a complete data matrix for 1 mitochondrial and 9 nuclear loci
. Sequences were aligned using Clustal X v.2.0.10 \citep{Larkin:2007hz}. We used PartitionFinder v.1.1.0 \citep{Lanfear:2012ei} to determine the best fit substitution models and partitioning schemes for each locus (See Supplementary Material, Table 1) and estimated individual gene trees using Bayesian phylogenetic analyses implemented in Beast v.1.8.0 \citep{beastlatest} with a strict molecular clock and a speciation yule-process tree prior. We obtained posterior distributions from one independent Markov chain Monte Carlo simulation, 
and assessed convergence with Tracer v.1.5  \citep{rambaut2013tracer}. After initial burn in of 100,000 trees, we took 100 samples from each posterior sampled at intervals of 100,000 trees. In this way we generated 10 gene trees $\times$  100 tree estimates on the same 10 species: $\{\hat{T}_j^{(i)}\}_{\{i = 1, \ldots, 10, j = 1,\ldots, 100\}}$ where $i$ indexes over the 10 genes and $j$ indexes over the posterior trees. We choose this level of thinning in order to generate a manageable number of near-independent draws from the posterior.

\begin{figure}
\begin{center}
\includegraphics[trim = 0cm 0.5cm 0.7cm 0.2cm, scale=0.9]{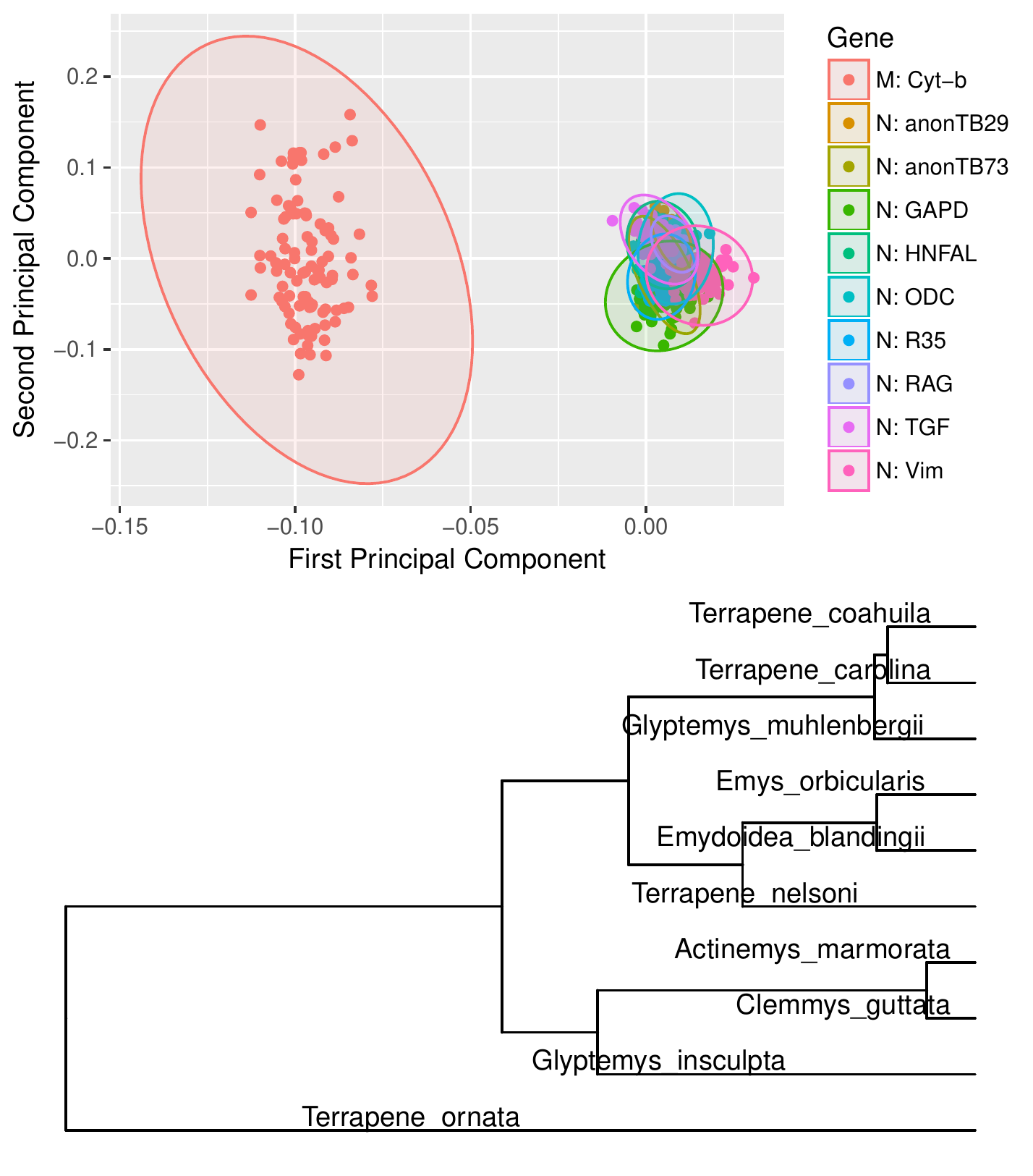}
\end{center}
\caption{(top) The 7-dimensional sets reflecting the variability in estimating the phylogenies of 9 nuclear (N) and 1  mitochondrial (M) gene in 10 species of Emydid  turtles.  The projection onto the first two  principal components, which explain 79\% of the variance, of the estimates is shown (see Supplementary Data for projections onto the third and fourth principal components).  We see that  the difference between the nuclear and mitochondrial trees are large relative to the within-phylogeny estimation error. (bottom) One of the estimates of the Vim gene tree.}
\label{tp}
\end{figure}

As described in Section \ref{intrinsic1}, we construct sets for the gene trees in $\mathbbm{R}^7$,  before projecting them onto first 2 principal components of the $\{\hat{T}_j^{(i)}\}_{i,j}$ (Figure \ref{tp}). Each gene tree generates its own covariance in accordance with the collection of estimates corresponding to that gene, for this reason the sets are not spherical nor self-similar.

We indeed notice discordance between the 1 mitochondrial ({\it cytochrome b }gene) and the 9 nuclear genes, but no strong disagreement between the nuclear gene phylogenies. Most importantly, the discordance between mitochondrial and nuclear loci is present even after accounting for the uncertainty in estimating the gene trees. It is possible that the uncertainty in estimating such a high-dimensional object (a tree in $\mathcal{T}_{10}$) could have swamped the apparent differences between the mitochondrial and nuclear phylogenies, as was the case in Section \ref{univariate}. We advocate for the statistical paradigm that a difference between groups is not large unless it is large relative to estimation error.

\section{Contrasting MDS with the log map} \label{contrast}

While the two procedures proposed in this paper have the advantage of a probabilistic interpretation that lend themselves to visualization of high-dimensional uncertainty, multidimensional scaling also has its advantages. For example, our procedure enforces that the distance between each tree estimate and the average tree (Fr\'echet mean) is measured according to the BHV distance. In contrast,  the choice of distance between trees in MDS can be selected to highlight the differing features between trees that are of greatest interest \citep{hillis2005analysis,Chakerian:2012kj,Kendall24062016}. In contexts where representing distances between trees is more important than displaying their estimation error, MDS will be a superior tool.

However, MDS has a number of serious drawbacks as a visualization method, some of which are not shared by the log map. We briefly describe five drawbacks of MDS that are improved by using the log map for visualization.

{\bf Visualizing uncertainty:}   MDS is a mapping rather than a rotation or projection, with the result that absolute  measures of uncertainties in tree estimates cannot be preserved. In contrast, the same projection applied to the log maps can be applied to a set, and thus absolute sizes of uncertainties can be reflected (such as in Figures \ref{orthomam2} and \ref{tp}).

\begin{center}
\begin{figure}
\includegraphics[trim = 0cm 1cm 1cm 0.5cm, scale=0.65]{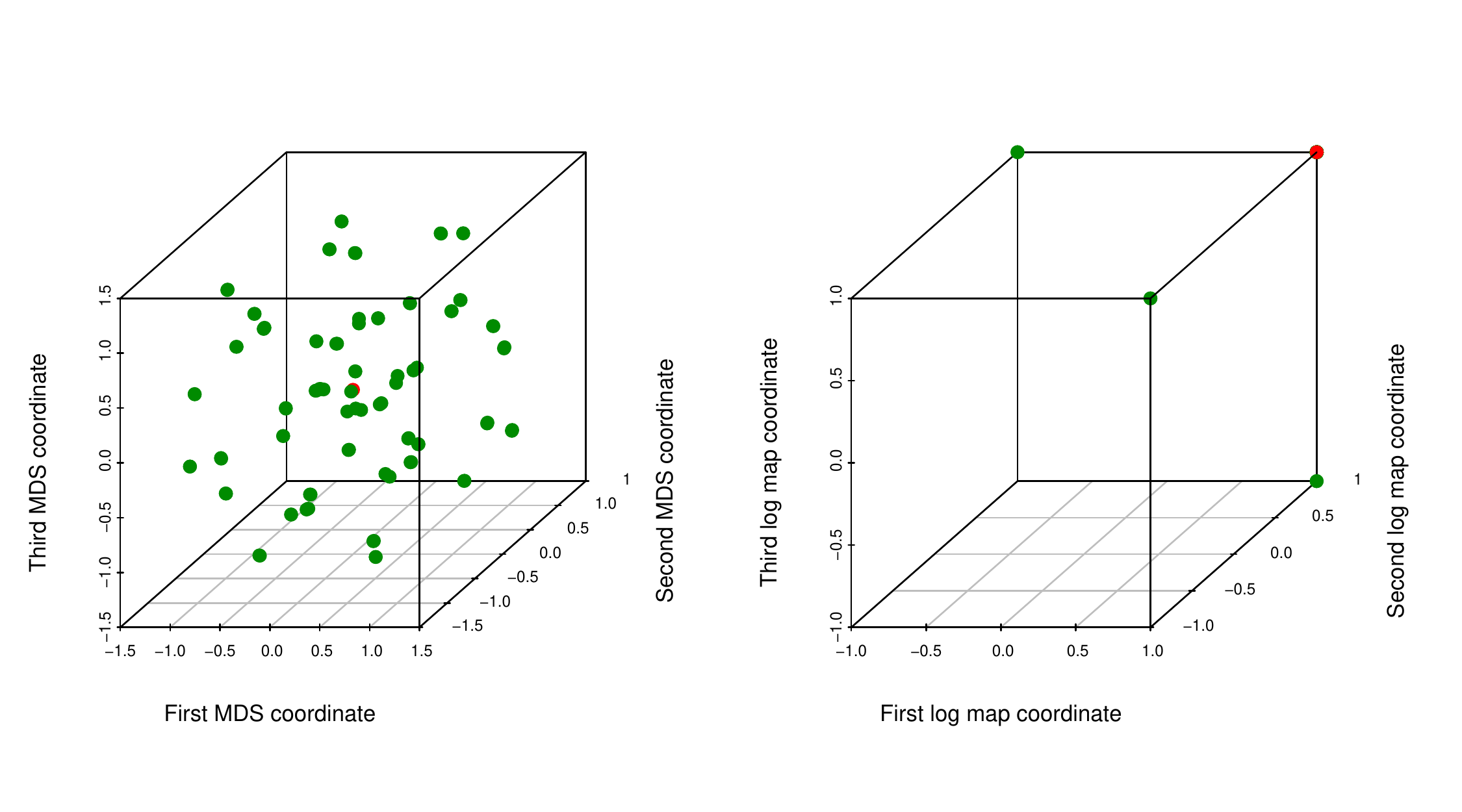}
\caption{Multidimensional scaling may distort visualization of equidistant trees. Visualization of NNI trees (green) from a 50-taxon tree (red) under MDS on BHV distances  using the Kruskal-1 stress function (left) and the   log map (right). All trees are equidistant from the base tree. MDS  distorts this, but the log map compresses trees onto one another.}
\label{distance_preserving}
\end{figure}
\end{center}

{\bf Distortions surrounding equidistance:} An ideal visualization procedure would communicate clearly which trees are equally distant from a central tree. Unfortunately, MDS fails to do this in many situations. To illustrate, we uniformly at random select a fully resolved tree topology with 50 leaves, and assign all branch lengths to unit length, and designate this tree as our ``base tree.'' We then consider the collection of all trees that are one nearest neighbor interchange (NNI) from this tree. According to both the RF and BHV distance, all trees are distance 2 from the base tree. However, the projection to 3 dimensions by MDS using the Kruskal-1 stress function (Eqn. (\ref{mds-equation})) would suggest that some trees are substantially closer than others (Figure \ref{distance_preserving}, left, modified from  \citet[Figure 10]{hillis2005analysis}). This can be very misleading when trying to identify representative trees or outliers. The log map, by construction, preserves distances to the base tree (Figure \ref{distance_preserving}, right), though the compression of multiple topologies onto a single point is sacrificed.

{\bf Reproducibility:} The stress-minimizing vector in MDS depends on every tree in the sample. As a result, a single new tree added to the sample may completely change the projection of all other points. Furthermore, visualizations cannot be compared across different studies, making the method inherently unreproducible.  As reproducibility becomes an increasing focus of genetic studies, the importance of reproducible figures  and visualization-based results is no less than reproducible quantitative analyses. As long as the base tree $\overline{T}$ and the PCA rotation matrices are maintained, the results of a new study can be compared alongside those of an original study {\it ex post facto}.

\begin{figure}
\begin{center}
\includegraphics[trim = 0cm 0cm 0cm 0cm, scale=0.7]
{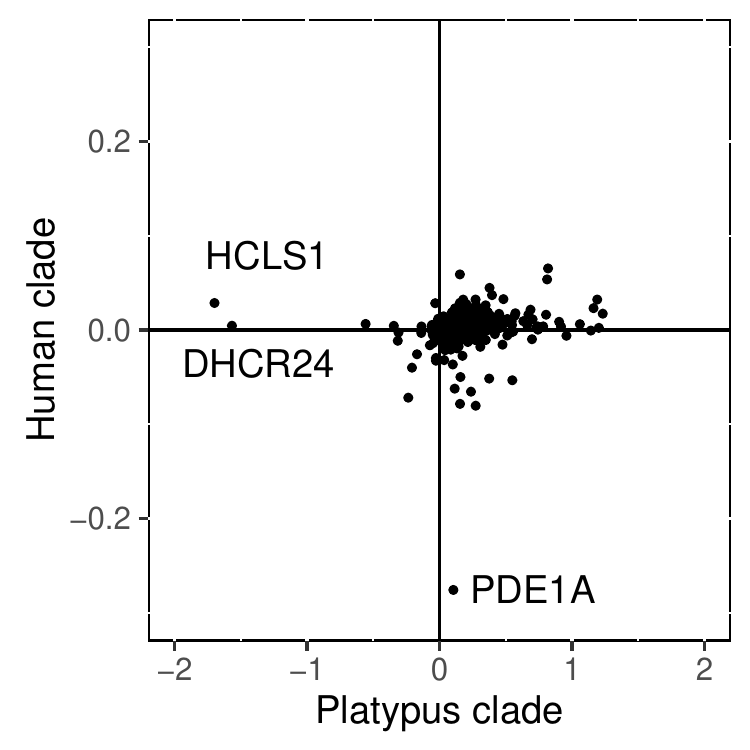}
\caption{Two coordinates of the log-mapped OrthoMam gene trees. A negative coordinate in a log map indicates that the branch is absent on the tree, while a positive coordinate indicates its presence. In this way the log map can distinguish trees that are topologically distinct from trees that have only different branch lengths, which is not possible using MDS on BHV distances. In this figure, the x-coordinate indicates the length of the branch separating the platypus from other marsupials, and the y-coordinate indicates the length of the branch separating the Human-Chimp-Gorilla clade from the remaining mammals. We give the names of three genes that are highly discordant on these branches.}
\label{mammals_branches}
\end{center}
\end{figure}

{\bf Topology versus branch length information:} Negative coordinates in a log map indicate that the topology of the tree is different to that of the base tree. A single negative coordinate in a log map indicates that the target tree is a nearest neighbor interchange from the base tree. Thus the log map is capable of distinguishing topological versus branch length differences. This information is lost under MDS (on non-RF distances). Furthermore,  if there is particular interest in a certain branch of the Fr\'echet mean tree, this coordinate could be plotted in order to investigate if a particular set of models or genes characterize the presence of the branch. In Figure \ref{mammals_branches}, we show two coordinates (branches) of the log map projection for the OrthoMam trees. The x-coordinate indicates the length of the branch separating the platypus from other marsupials (supported by 92\% of trees), while the y-coordinate indicates the length of the branch separating the Human-Chimp-Gorilla clade from the remaining mammals (supported by 81\% of trees). This information may be relevant to determining which clades on the mean tree are also supported by different genes.

{\bf Speed:}  Multidimensional scaling necessitates construction of the matrix of pairwise distances between trees, or calculation of  $\frac{n \times (n-1)}{2}$ distances. This can be  computationally prohibitive when distance calculations are expensive, as is the case for BHV distances. For example, in Section \ref{univariate}, constructing the BHV distance matrix for MDS required finding $\frac{574 \times 573}{2} = 164, 451$ geodesics. The procedure that we proposed required only 10,000 geodesics to update  $\bar{T}$ until convergence, followed by 574  geodesics to calculate the $\{\Phi_{\overline{T}}(\hat{T}_i)\}_i$. In practice, using the implementation of the geodesic calculation by \cite{nye11}, this amounted to 16 hours of computation for MDS compared to 5 minutes for our log map procedure, because the geodesics calculated using Algorithm 4.2 become progressively shorter, reducing the computational intensity of successive  geodesic calculations. Note that the more efficient mean calculation algorithms of \cite{Skwerer:H1QO6bI2} could increase the relative computational gains of our method, while a dynamic geodesic algorithm \citep{skwerer2015dynamic} could be used to reduce the computation time of MDS.

Nonetheless, we emphasize that the flexibility with respect to the metric distance of interest, as well as the preservation of relative distances between trees, maintain MDS as a valuable visualization tool, and we encourage its concurrent use even in situations where the model-based approach  proposed here may be useful.

\section{Limitations and open problems} \label{limitations}

As discussed in Section \ref{logmap},  the log map does compress tree space, and it is not a bijection. Thus if a collection of trees was uniformly distributed throughout tree space, the log map may dissolve the most important information about the collection. However, in practice, collections of trees are rarely distributed uniformly throughout tree space, and a small number of topologies usually characterize most trees in the sample. For this reason, dimension reduction via the log map can preserve much of the structure present in the collection of trees.  Indeed, this paper was motivated by attempts to find low rank approximations to tree collections. Nevertheless, unstructured tree datasets will be poorly reflected by the procedure proposed here.

Another criticism of the log map is that trees whose geodesic from the base tree follows the ``cone path'' will be mapped onto the line connecting the base tree to the star tree.  However, only 23\% of the Terrapene trees were a cone path from their mean tree, and zero of the 574 OrthoMam trees were a cone path from their mean tree. Furthermore, even if a large number of trees are cone path trees, this information can be used to determine which trees are highly topologically discordant. For example, 63\% of tree estimates for the ODC gene (Section \ref{intrinsic}) are cone path trees.  Thus we argue that, contrary to distorting the space, the log map is a useful tool for quickly diagnosing extreme topological discordance in certain trees.

A curiosity of the models  that were constructed in this paper  is that they model the difference  between tree estimates and true trees in Euclidean space, under the log map. It is more natural to model this difference in tree space, though unfortunately  development of probability distributions on tree space is in its infancy. Recent work showing convergence of random walks to Brownian motion in tree space provide a promising avenue for construction of probability distributions on tree space \citep{nye2015convergence}. However, Brownian motion in tree space is (locally) spherical,  and in its present form cannot describe non-spherical multivariate uncertainty in tree estimates. A promising direction of future research is the utility of probabilistic models {\it in tree space} for describing tree uncertainty, and comparisons with the models presented here.

A major limitation of almost all methods utilizing tree space is that missing leaves on some gene trees precludes their inclusion in the analysis. The procedure proposed here is no exception. While modern phylogenetic and phylogenomic data collection approaches (such as exon capture) are steadily reducing the amount of missing data, this remains a limitation.

\section{Concluding remarks} \label{conclusion}

We have proposed a method for incorporating tree-valued and non-tree-valued information  into visualizations of trees and their uncertainty. Multidimensional scaling has many advantages for tree visualization, but because the representation only exists in the context of  the sample, graphically incorporating the uncertainty that is {\it lost} in the map is not possible. By utilizing the log map to project the collection of trees to Euclidean space and modeling  the projections as noisy realizations of the underlying evolutionary process, we may scrutinize differences between tree estimates alongside their variabilities. The result is an interpretable, exploratory method for observing discordance between gene trees,  and among tree estimates. Further advantages of using PCA of log maps over multidimensional scaling include reproducibility and speed. We hope that the procedure and statistical perspective outlined in this paper reminds mathematicians working on tree space that trees are almost always estimated, and that differing levels of uncertainty needs to be accounted for when using tree space for analysis. However, most importantly, we hope to provide biologists with a method for diagnosing whether differences between gene trees are biologically meaningful, or due to uncertainty in estimation.

\section*{Acknowledgements}
The authors are very grateful to an anonymous referee and two editors, whose helpful and constructive suggestions substantially improved both the text and figures of the manuscript. We are also grateful to the \cite{rcore} and authors of the packages phangorn \citep{phangorn}, ape \citep{ape}, distory \citep{distory}, treespace \citep{treespace}, MASS \citep{mass}, scatterplot3d \citep{scatterplot}, XML \citep{xml}, gridExtra \citep{gridextra}, lattice \citep{lattice}, and ggplot2 \citep{ggplot}, which were used for constructing the figures and running the analyses in this paper.

\section*{Supplementary Materials and Data}
All data and scripts utilized in this paper are available at \\ \url{https://github.com/adw96/TreeUncertainty}. Details regarding model selection for Section \ref{intrinsic} and projecting ellipses onto principal components are available in the same location.

\bibliographystyle{Chicago}

\bibliography{all_references}
\end{document}